\documentclass[pra,letterpaper,aps,10pt,superscriptaddress,twocolumn,floatfix,showpacs]{revtex4-1}
\usepackage{graphicx}
\usepackage{amsmath}
\usepackage{amsfonts}
\usepackage{amssymb}
\usepackage{epsfig}
\usepackage{epstopdf}
\DeclareGraphicsExtensions{.pdf,.eps,.png,.jpg,.mps}
\usepackage[pdftex]{color}
\usepackage{amsmath,graphicx,amssymb,braket,xcolor,subfigure,upgreek}
\usepackage[colorlinks, linkcolor=blue, citecolor=blue, urlcolor=blue, breaklinks=true]{hyperref}
\usepackage{microtype}
\usepackage{bbm}
\usepackage{color}
\usepackage{braket}
\usepackage{amsmath,amssymb,amsfonts,latexsym,color,dcolumn,bm,amsbsy}
\usepackage[english]{babel}
\usepackage{graphicx}
\usepackage[utf8]{inputenc}
\usepackage{textcomp}
\usepackage{braket}

\newcommand{\omfb}{\omega_{\text{fb}}}
\begin{document}

\title{Partial Optomechanical Refrigeration via Multimode Cold-Damping Feedback}
\author{Christian Sommer}
\affiliation{Max Planck Institute for the Science of Light, Staudtstra{\ss}e 2,
D-91058 Erlangen, Germany}
\author{Claudiu Genes}
\affiliation{Max Planck Institute for the Science of Light, Staudtstra{\ss}e 2,
D-91058 Erlangen, Germany}
\affiliation{Department of Physics, University of Erlangen-Nuremberg, Staudtstra{\ss}e 2,
D-91058 Erlangen, Germany}
\date{\today}

\begin{abstract}
We provide a fully analytical treatment for the partial refrigeration of the thermal motion of a quantum mechanical resonator under the action of feedback. As opposed to standard cavity optomechanics where the aim is to isolate and cool a single mechanical mode, the aim here is to extract the thermal energy from many vibrational modes within a large frequency bandwidth. We consider a standard cold-damping technique where homodyne read-out of the cavity output field is fed into a feedback loop that provides a cooling action directly applied on the mechanical resonator. Analytical and numerical results predict that low final occupancies are achievable independently of the number of modes addressed by the feedback as long as the cooling rate is smaller than the intermode frequency separation. For resonators exhibiting a few nearly degenerate pairs of modes cooling is less efficient and a weak dependence on the number of modes is obtained. These scalings hint towards the design of frequency resolved mechanical resonators where efficient refrigeration is possible via simultaneous cold-damping feedback.
\end{abstract}

\maketitle

\begin{figure}[t]
\includegraphics[width=0.70\columnwidth]{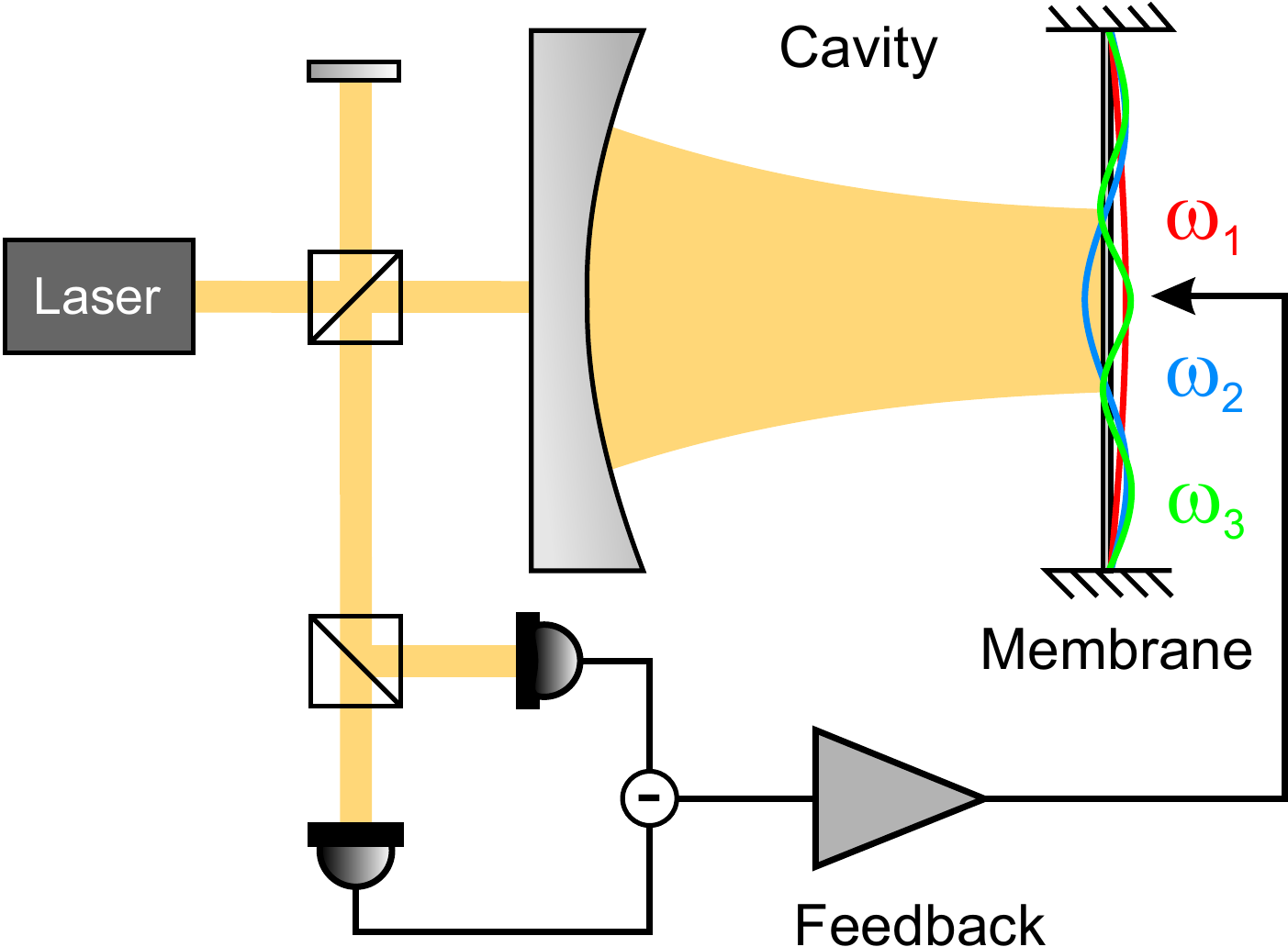}
\caption{\emph{Multi-mode cold damping}. The output of a driven optomechanical system is homodyne detected providing information on the collective displacement of many vibrational modes (\textit{bright mode}). A feedback loop is utilized to design a direct force to be applied onto the resonator that can lead the freezing of its thermal fluctuations (cold-damping effect). }
\label{fig1}
\end{figure}

In recent decades great progress has been accomplished in laser cooling of microscopic and macroscopic objects, ranging from atoms, ions and molecular systems to selected modes of micro-mechanical oscillators or levitated nanoparticles~\cite{Metcalf1999Laser,Aspelmeyer2014cavity}. In standard cavity quantum optomechanics with macroscopic mechanical resonators or levitated nanoparticles a crucial goal is to isolate and cool a given vibrational mode of interest~\cite{Aspelmeyer2014cavity,windey2019cavity,deli2019cavity, Rossi2017Enhancing, Clark2017Sideband, Qiu2019High,Asenbaum2013Cavity,Mancini1998Optomechanical, Clemens2016Quantum, Kiesel2013Cavity,Millen2015Cavity,rodenburg2016quantum,khosla2017quantum} close to its quantum ground state. This can for example be utilized towards high-precision sensing applications. Different techniques have been employed among which two stand out: cavity resolved sideband cooling (or cavity-assisted cooling)~\cite{Gigan2006self, Braginsky2007parametric, Marquardt2007Quantum, Wilson2007Theory, Teufel2011Sideband} and feedback-aided cooling (in particular the cold-damping technique)~\cite{genes2008ground, steixner2005quantum, Bushev2006Feedback,rossi2018measurement, Cohadon1999Cooling, Poggio2007Feedback, Wilson2015Measurement,conangla2019optimal,tebbenjohanns2019cold}. As mechanical resonators typically exhibit a large number of vibrations, cooling of a single mode leaves the overall temperature of the object largely unaltered. For regimes where many vibrational modes can be found within a single cavity resonance it is interesting to ask what is the efficiency of cold-damping in the simultaneous reduction of the occupancy of a few modes. This could lead to the partial refrigeration of mechanical resonators with enhanced sensing capabilities in a much larger frequency bandwidth.\\
\indent We provide here a theoretical investigation of cold damping simultaneously applied to $N$ mechanical resonances~\cite{Nielsen2017Multimode,Piergentili2018Two,Wei2019Contollable} in order to provide a roadmap for partial refrigeration. The analytical treatment consists in finding solutions for a set of quantum Langevin equations describing the evolution of $N$ vibrational modes coupled to a single optical mode and to thermal and optical reservoirs. In this sense our approach is general and could be tailored to a variety of systems such as a set of nano-particles in dipole traps, or ions in an ion trap, where the thermal excitations is distributed into $N$ collective oscillation modes. However, for the simplicity of presentation we will only refer to the system depicted in Fig.~\ref{fig1} consisting of an optomechanical cavity with a movable end-mirror or highly-reflective membrane. The cavity output with frequency components corresponding to the radiation pressure coupling to many vibrational modes, is passed through a feedback device that allows for the choice of a correct back-action onto the mechanical resonator that can compensate the heating effect of the environment.
We analyze the final occupancies of all modes undergoing cold damping and compare this to the expected results obtained for cooling of independent isolated modes~\cite{genes2008ground}. As the readout field provides information solely on a \textit{bright mode} (the generalized quadrature that the cavity field couples to) the feedback only directly cools this mode. It would then be expected that the uncoupled $N-1$ orthogonal \textit{dark modes} would considerably slow down the overall refrigeration process. This would be indeed the case for completely degenerate mechanical modes where only $1/N$ of the total thermal energy can be removed. Instead, for non-degenerate modes, the main theoretical result of this calculation (backed by numerical simulations) indicates that the efficiency of the cooling process is roughly independent of the number of modes undergoing cooling dynamics. This occurs as inter-mode frequency mismatches offer dark to bright mode couplings leading to an efficient sympathetic cooling mechanism for all modes involved. A detrimental aspect of increasing the number of modes is the enhanced probability of finding consecutive modes of nearly degenerate frequency. The result previously pointed out in the case of resolved sideband cooling of two mechanical modes, holds here in the cold-damping case as well, which is that the cooling efficiency is degraded as modes approach the point of degeneracy~\cite{genes2008simultaneous,korppi2019sideband}. This indicates a roadmap to efficient partial refrigeration consisting of the design of mechanical resonators with close to linearly spaced normal mode frequencies where the optical cooling rate is only limited by the minimal inter-mode frequency separation.\\

\noindent \textbf{Linearized Langevin equations} --- We consider an optomechanical cavity with a movable end-mirror or highly-reflective membrane exhibiting $N$ independent modes of vibrations each of effective mass $m_j$ and frequency $\omega_j$ (as depicted in Fig.~\ref{fig1}). The quantum motion of the mechanical modes is described by the displacement $Q_j$ and momentum $P_j$ quadrature operators with standard commutations $[Q_j,P_{j'}]=i\hbar\delta_{jj'}$. A single cavity mode at frequency $\omega$ and loss rate $\kappa$ is described by a bosonic operator $A$ with $[A,A^\dagger]=1$. The optomechanical interaction is a standard radiation pressure Hamiltonian $\sum_j\hbar g_\text{OM}^{(j)} A^\dagger A Q_j$ where $g_\text{OM}^{(j)}$ is the single photon single phonon coupling rate. Driving is executed with a laser of power $\cal{P}$ and frequency $\omega_\ell$ via the non-moving mirror at rate $\epsilon = \sqrt{2{\cal{P}}\kappa/\hbar \omega_\ell}$ (where $\kappa$ is the photon loss rate). We follow a standard quantum Langevin treatment of optomechanics~\cite{genes2008ground} (see Appendix) where the operators are split into classical averages plus zero-average quantum fluctuations: $A=\braket{A}+a$, $Q_j=\braket{Q_j}+q_j$ and $P_j=\braket{P_j}+p_j$. The problem then becomes linear in the limit $|\braket{A}|\gg 1$ and the following set of Langevin equations can be written for the fluctuations:
\begin{subequations}
\label{CDEq.1}
\begin{align}
\dot{q}_j &= \omega_j p_j,\\
\label{CDEq.1a}
\dot{p}_j &= -\omega_j q_j - \gamma_j  p_j +G_j x- g_j*y^{\text{est}}+ \xi_j,\\
\dot{x} &= -\kappa x + \sqrt{2\kappa} x^\text{in},\\
\dot{y} &= -\kappa y + \textstyle \sum_{j=1}^{N} G_j q_j+ \sqrt{2\kappa} y^\text{in},
\end{align}
\end{subequations}
where $x = (a + a^{\dagger})/\sqrt{2}$ and $y = i(a^{\dagger} - a)/\sqrt{2}$ are the quadratures of the cavity field fluctuations and $x^{\text{in}}$, $y^{\text{in}}$ are the corresponding optical input noise terms similarly defined from the input optical noise operator $a^\text{in}$. The zero-average noise terms are delta-correlated in time $\braket{a^\text{in}(t)a^{\text{in}\dagger}(t')}=\delta(t-t')$. The effective optomechanical couplings $G_j=\sqrt{2}g_\text{OM}^{(j)}\braket{A}$ are enhanced by the large cavity field amplitude. We set the condition that the effective cavity detuning $\Delta=\omega-\omega_\ell-\textstyle \sum_j g_{\text{OM}}^{(j)} \braket{Q_j}$, containing a collective mechanically-induced frequency shift is kept at zero value (see Appendix). For each mode $j$ thermalization with the environment is described by a zero-averaged Gaussian stochastic noise term $\xi_j$ and a rate $\gamma_j$. The fluctuation-dissipation relation is fulfilled as described by the two-time correlation function
$\langle \xi_j(t) \xi_{j'}(t')\rangle = \gamma_{j}/\omega_j\int_{0}^{\Omega} d\omega/2\pi e^{-i\omega(t-t')} S_{\text{th}}(\omega)\delta_{jj'}$
where $\Omega$ is the frequency cutoff of the reservoir and $S_{\text{th}}(\omega)=\omega[\coth\left(\hbar \omega/2k_B T\right) + 1]$ is the thermal noise spectrum. The correlation function becomes a standard white noise input with delta correlations both in frequency and time for sufficiently high temperatures $k_B T \gg \hbar \omega_j$. This results in the approximate form $\langle \xi_j(t) \xi_{j'}(t')\rangle \approx (2\bar{n}_{j}+1)\gamma_{j}\delta(t-t')\delta_{jj'}$, where the occupancy of each vibrational mode is given by $\bar{n}_{j} = 1/\left[\exp(\hbar \omega_j/k_B T)-1\right] \approx k_B T/\hbar \omega_j$. \\
Notice that the cavity field only provides information on a generalized mode which we dub \textit{bright mode} obtained as a linear combination of individual modes quadratures. All other collective modes orthogonal to this one could then be defined as \textit{dark modes} and apparently do not participate in the cooling dynamics. However, as we will show, intermodal correlations are built-up during the dynamics and all individual modes are affected.\\

\noindent \textbf{Feedback loop} --- The feedback force on the $j$'s mode given by the convolution term $(g_j \ast y)(t)=\int_{-\infty}^{\infty}dsg_j(t-s)y(s)$ depends on past dynamics of the detected quadrature $y$ that is driven by the weighted sum of the oscillator fluctuations $q_j$. Here, the causal kernel
\begin{align}
\label{gcd}
g_j(t) =g_\text{cd}^{(j)} \partial_t \left[ \theta(t)\omfb e^{-\omfb t}\right]
\end{align}
contains the feedback gain terms $g_\text{cd}^{(j)}$ and feedback bandwidth $\omfb$. Notice that in the limit $\omfb \rightarrow \infty$ the feedback becomes $g_j(t) = g_\text{cd}^{(j)}\delta'(t)$. The component injected into the feedback loop $  y^{\text{est}}$ is the estimated intracavity phase quadrature. This results from a measurement of the output quadrature $y^{\text{out}} = \sqrt{2\kappa} y(t) - y^{\text{in}}(t)$ and additionally considering a detector with quantum efficiency $\eta$ (modeled by an ideal detector preceded by a beam splitter with transmissivity $\sqrt{\eta}$, which mixes the input field with an uncorrelated vacuum field $y^v(t)$). The estimated signal is then written as
\begin{align}
y^{\text{est}}(t) &= y(t) - \frac{y^{\text{in}}(t) + \sqrt{\eta^{-1} -1}y^v(t)}{\sqrt{2\kappa}}.
\end{align}

\noindent \textbf{Multi-mode cold damping} --- We proceed by first formally eliminating the dynamics of the cavity quadratures to find a set on $2N$ nonlinear differential equations for the dynamics of mechanical modes. Under the conditions of fast feedback and fast cavity dynamics, we then can simplify this to a set of linear differential equations with analytical solutions. We start by writing a formal solution for the $y$-quadrature as $y(t) = \int_{-\infty}^{t}ds e^{-\kappa(t-s)}\left(\textstyle \sum_{j=1}^{N}G_jq_j(s)+y^\text{in}(s)\right)$
and estimate the effect of the cold-damping convolution term appearing as a drive term for the membrane's momentum in Eq.~\ref{CDEq.1a} (see Appendix)
\begin{eqnarray}
\label{CDEq.3}
\nonumber
(g_j \ast   y)(t) = g_\text{cd}^{(j)} \omfb \textstyle \sum_{k=1}^{N}  G_k \omega_k (h\ast  p_k)(t)+\xi_{y^\text{in}}.
\end{eqnarray}
The right-hand side term contains the added cold-damping decay rate acting on the $j$'s mode as well as some cross-terms which dissipatively couple distinct modes. In the lossy cavity limit $\kappa \gg \omega_j$ and fast feedback $ \omfb  \gg \omega_j$ the first term of the convolution above can be approximated by an instantaneous value $p_j/(\kappa \omfb)$. This allows one to turn Eqs.~\ref{CDEq.1} into a set of linear differential equations:
\begin{subequations}
\begin{align}
\label{CDEq.5}
  \dot{q}_j &= \omega_j   p_j,\\
  \dot{p}_j &= -\omega_j   q_j - \Gamma_{jj}  p_j -\textstyle \sum_{k\neq j}\Gamma_{jk}p_k+ \xi_j+\xi_{\text{opt},j}.
\end{align}
\end{subequations}
\begin{figure*}[t]
\includegraphics[width=2.00\columnwidth]{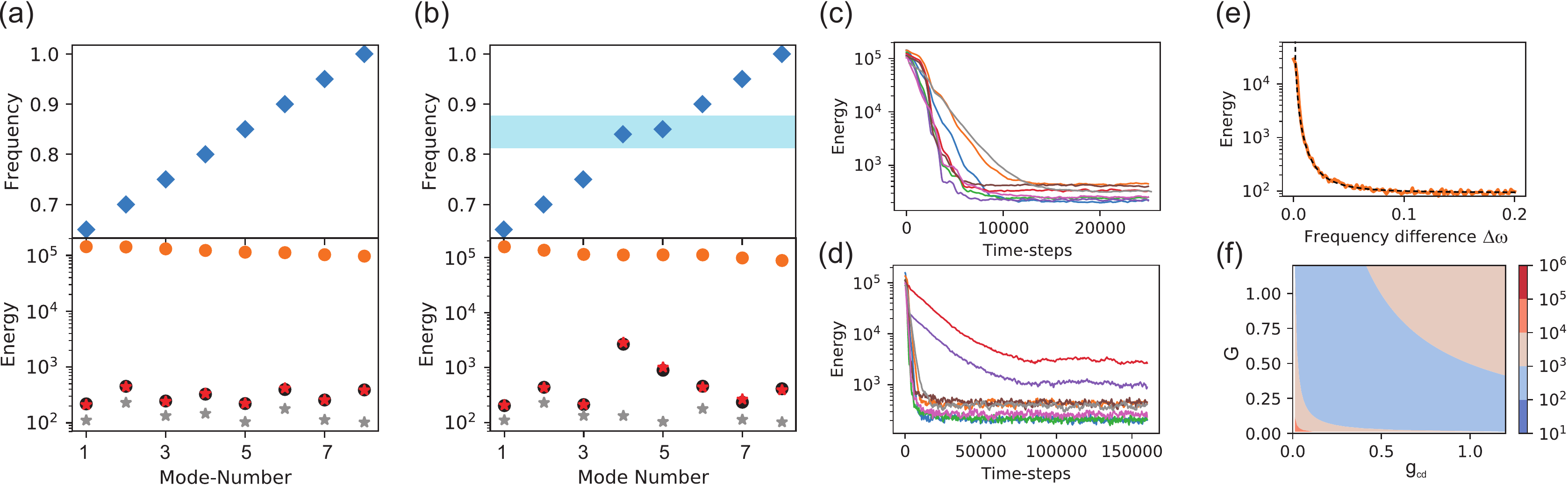}
\caption{\emph{Multimode cold damping}. Comparison between cooling efficiency in the case of linear dispersion  in (a) versus dispersion relation showing two quasi-degenerate modes in (b) for a resonator with eight independent vibrational modes. The initial (orange disks) and final occupancies (black disks), obtained from a Monte Carlo simulation, are displayed and compared to the predictions of the Lyapunov equation (red stars). Results obtained from a simplification assuming  independent damping of each oscillator are indicated by gray stars. For nearly-degenerate modes cooling is strongly inhibited while the other modes are virtually unaffected. In (c),(d) the average energies are presented as a function of time for case (a) in (c) and case (b) in (d). In (e) dependence of the occupation number normalized final energy with respect to the frequency difference between two modes is obtained from a Monte Carlo simulation (orange solid line) and via solving the Lyapunov equation (black dashed line). In (f) the final occupancy of an individual mode is plotted versus variations of the intra-cavity light amplitude (affecting all the $G_j$) and feedback gain.}
\label{fig2}
\end{figure*}
The added optical noise $\xi_{\text{opt},j}$ is a sum of a feedback induced noise term as well as the direct radiation pressure noise.
The diagonal term contains the cold-damping optical loss
\begin{equation}
\Gamma_{jj}=\gamma_j + \frac{g_\text{cd}^{(j)}G_j\omega_j}{\kappa},
\end{equation}
expected for a single isolated uncoupled mode~\cite{genes2008ground}. The presence of adjacent modes leads to a dissipative cross-talk at similar rates $\Gamma_{jk}=g_\text{cd}^{(j)} G_k\omega_k/\kappa$. The back-action both due to feedback as well as to radiation pressure effects is contained within the term $\xi_\text{opt}$ which adds to the thermal noise. In a simplified regime where all $\Gamma_{jk}=\Gamma$ the generalized quadrature $\sum_j p_j/\sqrt{N}$ defines the bright mode which is directly addressed by the feedback loop. For degenerate frequencies this is the only mode effectively cooled at a damping rate $N\Gamma$. All other $N-1$ collective modes orthogonal to the bright one become effectively dark - not directly addressed by the feedback loop. As the thermal energy is equally spread in all collective modes only $1/N$ of total energy of the resonator is extracted. In the non-degenerate case, frequency mismatches lead to coupling of dark modes to the bright mode and thus to a mechanism of efficient sympathetic cooling of all degrees of freedom present (see Appendix for the derivation of dark-bright mode couplings similar to the procedure applied in Ref.~\cite{delpino2019tensor}).\\
The $2N$ equations of motion can be cast in a compact form given by $\dot{\bold{v}} = M   \bold{v} + \bold{n}_{\text{in}}$ where $ \bold{v} = (  q_1,   p_1,   q_2,   p_2, \dots ,   q_N,   p_N )^{\top}$ and $\bold{n}_{\text{in}} = (0,\xi_1+\xi_{\text{opt},1}, \dots , 0, \xi_N+\xi_{\text{opt},N})^{\top}$. In the case that the solution is stable and all the eigenvalues of the matrix $M$ have negative real parts, the system achieves a steady state fully characterized by the covariance matrix $V = \langle   \bold{v}(t)   \bold{v}^{\top}(t)\rangle$. Under the assumption that the cavity is lossy and the feedback is fast one can then show that the diffusion matrix is delta-correlated $\langle \bold{n}_{\text{in}}(t) \bold{n}_{\text{in}}^{\top}(t')\rangle = {\cal{D}}_{\text{in}}\delta(t-t')$ and the covariance matrix is computed by solving a steady state Lyapunov equation $MV + VM^{\top} =-{\cal{D}}_{\text{in}}$. For $\Gamma_{jj} \approx  (g_\text{cd}^{(j)}G_j\omega_j)/\kappa $ one can then analytically estimate the momentum and displacement fluctuation variances as:
\begin{widetext}
\begin{subequations}
\begin{align}
\langle p_i^2\rangle &= \left(\bar{n}_i + \frac{1}{2} \right)\frac{\gamma_i}{\Gamma_{ii}} + \frac{G^2_i}{2\Gamma_{ii}\kappa} + \sum_{j \neq i}\frac{\Gamma_{ij}}{2\Gamma_{ii}}\left\{ \frac{\left(\omega_i^2\Gamma_{jj} + \omega_j^2\Gamma_{ii} \right)\Lambda_{ij}}{\left(\omega_i^2 -\omega_j^2 \right)^2} + \sum_{k \neq i,j} \frac{1}{\left( \omega_i^2 -\omega_j^2\right)}\left(\frac{\omega_i^2 \Gamma_{jk} \Lambda_{ik}}{\left(\omega_i^2 -\omega_k^2 \right)} - \frac{\omega_j^2 \Gamma_{ik} \Lambda_{jk}}{\left(\omega_j^2 - \omega_k^2 \right)} \right) \right\}, \\
\langle q_i^2\rangle &= \langle p_i^2\rangle + \sum_{j \neq i} \frac{\Gamma_{ij}\Lambda_{ij}}{2\left(\omega_i^2 - \omega_j^2 \right)},
\end{align}
\end{subequations}
\end{widetext}
where we have defined the following quantity
\begin{widetext}
\begin{align}
\Lambda_{ij} &:=\frac{g^{(j)}_{\text{cd}}}{g^{(i)}_{\text{cd}}}(2\bar{n}_i + 1)\gamma_i + \frac{g^{(i)}_{\text{cd}}}{g^{(j)}_{\text{cd}}}(2\bar{n}_j + 1)\gamma_j + \frac{(g^{(j)}_{\text{cd}}G_i - g^{(i)}_{\text{cd}}G_j)^2}{\kappa g^{(i)}_{\text{cd}}g^{(j)}_{\text{cd}}}.
\end{align}
\end{widetext}
Notice that the apparent divergence in the denominators comes from neglecting the bare mechanical damping rate $\gamma_i$ with respect to any of the diagonal or mutual damping rates $\Gamma_{ij}$. The regime of interest where we compare analytical results with numerical simulations assumes non-degenerate modes where the frequency separation is much larger than any $\gamma_i$. Exact results (a bit more cumbersome) can however be obtained for any case as shown in the Appendix. \\

\noindent \textbf{Discussions} --- A few observations can immediately be made on the expressions above. First, one notices that the equipartition theorem is generally not fulfilled signifying that the final state cannot be described by an effective temperature. Then, in the absence of mode-mode coupling (where all $\Gamma_{ij}$ are set to zero) the result is the expected one~\cite{genes2008ground} where the final state fulfills the equipartition theorem and achieves an occupancy of roughly $\bar{n}_i \gamma_i/\Gamma_{ii}$. Notice that there is also a residual occupancy coming from the noise in the $x$ quadrature; this can be neglected under the assumption that the cavity-assisted cooling rate $G_i^2/2\kappa$ is much smaller than the feedback damping rate $\Gamma_{ii}$ easily fulfilled when the gain is large enough such that $g^{(i)}_{\text{cd}}\omega_i\gg G_i$. Extra terms arise from inter-mode momentum-momentum $\braket{p_i p_j}$ and momentum-displacement $\braket{p_i q_j}$ quantum correlations due to the fact that the feedback force term contains a sum of all momentum quadratures.\\
In the limit of linear dispersion relation where $\omega_j \approx \omega + j\Delta \omega$ and $\omega \gg \Delta \omega$ the final occupancy can be easily simplified to (see Appendix)
\begin{eqnarray}
\label{Eq:Ei}
\nonumber
{\cal{E}}_i &\approx & \bar{n}_i\frac{\gamma_i}{\Gamma_{ii}} + \frac{1}{4\Delta \omega^2}\sum_{\begin{smallmatrix} \langle i,j \rangle \\ j \neq i \end{smallmatrix}}\frac{\Gamma_{ij}}{2}\left\{ \left(1 + \frac{\Gamma_{jj}}{\Gamma_{ii}}\right)\Lambda_{ij} \right. \\\nonumber
& & + \left. \sum_{\begin{smallmatrix} \langle i,j,k \rangle \\ k \neq i,j \end{smallmatrix}} \frac{1}{\Gamma_{ii}}\left(\frac{\Gamma_{jk}\Lambda_{ik}}{(i-j)(i-k)}-\frac{\Gamma_{ik}\Lambda_{jk}}{(i-j)(j-k)} \right) \bigg\}  \right., \\
\end{eqnarray}


where we have defined the energy per mode as ${\cal{E}}_i=(\langle p_{i}^2\rangle+\langle q_{i}^2\rangle)/2$ and where we can approximate $\Lambda_{ij} \approx (2\bar{n}_i + 1)\gamma_i + (2\bar{n}_j + 1)\gamma_j$. Notice that the sums are only performed on the nearest neighbors in frequency space. This is a crucial aspect as one sees that the heating effects stemming from the coupling to the neighbor modes do not scale with the number of modes $N$ but only depend on the relative inter-mode minimal distance. For the case that $\Gamma_{ii},\Gamma_{ij}\ll \Delta\omega$ the expression above indicates that the final occupancies are close to what would have been expected from $N$ independent feedback loops, each specialized to a given mode.\\
The comparison between numerical and analytical results in Fig.~\ref{fig2}a,b indicate that nearly-degenerate modes are cooled with a lower efficiency. In fact, the expression above reproduces well this effect showing that the maximum optical cooling rate cannot exceed the inter-mode frequency separation $\Delta \omega$. To further elucidate this aspect, we focus in Fig.~\ref{fig2}e on a target mode at frequency $\omega$ and plot its final occupancy in the presence of an adjacent mode at frequency $\omega+\Delta\omega$ as a function of the frequency separation between the two modes. The final occupancy of mode $\omega$ is unaffected for $\Delta \omega >\Gamma$ while at near degeneracy the cooling is completely inefficient. In the intermediate regime, the scaling with $(\Delta \omega)^{-2}$ predicted by Eq.~\ref{Eq:Ei} holds.\\
While the damping rate includes a product of the cavity intra-cavity field amplitude (via the coefficients $G_j$) with the feedback gain (via the terms $g_\text{cd}^{(j)}$), back-action noise limits the final achievable occupancy. Figure.~\ref{fig2}f shows a density plot of the final occupancy of a target mode as a function of these two parameters. We have kept the relative ratios of all the parameters $G_j$ and $g_\text{cd}^{(j)}$ and simultaneously varied them. The result shows that past the optimal regime for cooling the radiation pressure and feedback noise add to the achievable final occupancy.\\

\noindent \textbf{Conclusions}---We have theoretically analyzed prospects for using a single cold-damping feedback loop for the partial refrigeration of a multi-mode mechanical resonator within a large frequency window. As a main result we have derived an analytical expression for the final occupancy of the modes undergoing cooling dynamics. The expression predicts a generalization of a previously known result obtained in the case of two mechanical modes, i.e. that efficient refrigeration requires the absence of nearly-degenerate vibrations in the mechanical spectrum of the resonator. This can be understood as a sympathetic cooling mechanism where a bright collective mode is directly damped by the feedback loop and provides sympathetic cooling of dark modes to which is coupled via frequency disorder. Under these conditions, simultaneous cooling can achieve final temperatures close to the case of isolated mode addressing.\\
\indent While we have illustrated our approach on the many modes of a macroscopic mechanical resonator, our analysis is quite general as it can also be applied to the cooling of collective vibrational modes of ions in ion traps, of atoms in dipole traps, levitated nanoparticles in optical tweezers etc. For interacting systems held in the same externally designed potential (as is the case of ion traps), the frequency spectrum of the collective modes can be tailored to eliminate near degeneracy points and single particle addressing ensures a simultaneous driving of all collective modes. For mesoscopic systems exhibiting a limited number of mechanical resonances the mechanism can lead to the full refrigeration of the object's motion. This could be for example achieved in two-dimensional atomic array optomechanics~\cite{shahmoon2018quantum,shahmoon2019collective} or in ensembles of interacting levitated nanoparticles in neighboring optical tweezer traps.\\
\indent Finally, let us comment on the assumed form of cold-damping feedback: in Eq.~\eqref{gcd} we have assumed that the electronic loop can provide an instantaneous feedback onto the system. This assumption is contained in the argument of the Heaviside function $\theta(t)$. This assumes fast electronics which can respond much quicker than the oscillation time of the system. For some current experiments~\cite{rossi2018measurement} feedback delay is an issue as it can lead to inadvertent heating of the target mode instead of the envisioned cooling~\cite{zippilli2018cavity}. The effect of a time delay $\tau_{\text{fb}}$ can be included in the argument $\theta(t-\tau_{\text{fb}})$ and will be addressed in the future following a similar formalism.\\

\noindent\textbf{Acknowledgments -} We acknowledge financial support from the Max Planck Society and from the German Federal Ministry of Education and Research, co-funded by the European Commission (project RouTe), project number 13N14839 within the research program "Photonik Forschung Deutschland". We acknowledge initial discussions with Muhammad Asjad and very useful comments on the manuscript from Andr\'{e} Xuereb and David Vitali.

\bibliography{ColdDamping}

\newpage
\onecolumngrid

\newpage
\appendix

\section{Linearized quantum Langevin equations in optomechanics}
\label{A}

We start with a quantum formulation of the system's dynamics of a few independent oscillation modes with frequencies $\omega_j$ of a membrane resonator (where $j=1,...N$). The equations of motion for the collection of modes are written as
\begin{subequations}
\begin{align}
\label{App.ModEq.1}
\dot{Q}_j &= \omega_j P_j, \\
\dot{P}_j &= -\omega_j Q_j  -\gamma_j P_j + g_{\text{OM}}^{(j)} A^{\dagger}A + \xi_j,\\
\dot{A} &= -(\kappa + i\Delta_0)A + i\textstyle \sum_{j=1}^{N} g_{\text{OM}}^{(j)} A Q_j + \epsilon + \sqrt{2\kappa}a^{\text{in}},
\end{align}
\end{subequations}
in terms of dimensionless position and momentum quadratures $Q_{j}$ and $P_{j}$ for each of the $N$ independent membrane oscillation modes. The term $\Delta_0 = \omega_c-\omega_\ell$ describes the detuning of the cavity resonance frequency $\omega_c$ from the laser frequency $\omega_l$ and $\kappa$ its decay rate. The oscillator frequencies are given by $\omega_{j}$. The radiation pressure coupling is given by $g_{\text{OM}}^{(j)}$ for the $j$'s mode and the input laser power by $\epsilon = \sqrt{2\mathcal{P}\kappa/\hbar \omega_\ell}$. The damping of the $j$'s resonator mode is described by the parameter $\gamma_{j}$ and is with the associated zero-averaged Gaussian stochastic noise term leading to thermalization with the environment. The noise term can be fully described by the two-time correlation function:
\begin{eqnarray}
\label{App.ModEq.2}
\langle \xi_j(t) \xi_{j'}(t')\rangle &=& \frac{\gamma_{j}}{\omega_j}\int_{0}^{\Omega} \frac{d\omega}{2\pi}e^{-i\omega(t-t')} S_{\text{th}}(\omega)\delta_{jj'},
\end{eqnarray}
where $\Omega$ is the frequency cutoff of the reservoir and $S_{\text{th}}(\omega)=\omega[\coth\left(\hbar \omega/2k_B T\right) + 1]$ is the thermal noise spectrum. The correlation function becomes a standard white noise input with delta correlations both in frequency and time for sufficiently high temperatures $k_B T \gg \hbar \omega_j$. This results in the approximate form $\langle \xi_j(t) \xi_{j'}(t')\rangle \approx (2\bar{n}_{j}+1)\gamma_{j}\delta(t-t')\delta_{jj'}$, where the occupancy of each vibrational mode is given by $\bar{n}_{j} = (\exp(\hbar \omega_j/k_B T)-1)^{-1} \approx k_B T/\hbar \omega_j$.
The cavity input noise is described by $a^{\text{in}}$ and follows the the correlation function $\langle a^{\text{in}}(t) a^{\dagger\text{in}}(t')\rangle = \delta(t-t')$. For an intense cavity field and rewriting the operators $A = \langle A \rangle + a$, $Q_j = \langle Q_j \rangle + q_j$ and $P_j = \langle P_j \rangle + p_j$ as a sum of their expectation value and a fluctuation term we can simplify the equations of motion in Eq.~\ref{App.ModEq.1} and obtain for the equations of motion for the expectation values
\begin{subequations}
\begin{align}
\label{App.ModEq.3}
\langle \dot{Q}_j \rangle &= \omega_{j} \langle P_{j}\rangle,\\
\langle \dot{P}_j \rangle &= -\omega_{j} \langle Q_{j} \rangle - \gamma_j \langle P_j \rangle + g_{\text{OM}}^{(j)} |\langle A \rangle|^2,\\
\langle \dot{A}\rangle &= -(\kappa + i\Delta_0)\langle A \rangle + i\sum_{j=1}^{N} g_{\text{OM}}^{(j)} \langle A \rangle \langle Q_j \rangle + \epsilon,
\end{align}
\end{subequations}
which at steady ($\langle \dot{Q}_j \rangle = \langle \dot{P}_j \rangle = \langle \dot{A} \rangle = 0$) state results in
\begin{eqnarray}
\label{App.ModEq.4}
\langle A \rangle &=& \frac{\epsilon}{\left[\kappa + i\left(\Delta_0 - \sum_{j} \frac{\left( g_{\text{OM}}^{(j)} \right)^2}{\omega_j}|\langle A \rangle|^2 \right) \right]},
\end{eqnarray}
where $\Delta = \Delta_0 - \sum_{j} \frac{\left( g_{\text{OM}}^{(j)}\right)^2}{\omega_j}|\langle A \rangle|^2$ is the effective cavity detuning including radiation pressure effects and $\langle Q_j \rangle = (g_{\text{OM}}^{(j)}/\omega_j)|\langle A \rangle|^2$. \\
For the fluctuations where we can omit all nonlinear terms $a^{\dagger}a$ and $a q_j$ since $|\langle A \rangle| \gg 1$, we obtain the linearized equations of motion
\begin{subequations}
\begin{align}
\label{App.ModEq.5}
\dot{q}_j &= \omega_j p_j,\\
\dot{p}_j &= -\omega_j q_j - \gamma_j p_j + \xi_j + G_j x,\\
\dot{x} &= -\kappa x + \Delta y + \sqrt{2\kappa} x^{\text{in}}, \\
\dot{y} &= -\kappa y - \Delta x + \textstyle \sum_{j=1}^{N} G_jq_j + \sqrt{2\kappa}y^{\text{in}},
\end{align}
\end{subequations}
where $x = (1/\sqrt{2})(a+ a^{\dagger})$ and $y = (i/\sqrt{2})(a^{\dagger}-a)$ are the quadratures of the cavity field and $x^{\text{in}}$ and $y^{\text{in}}$ are formulated correspondingly. The effective optomechanical coupling terms are given by $G_j = \sqrt{2}g_{\text{OM}}^{(j)}\langle A \rangle$.\\

\section{Multimode cold damping}
\label{B}

For cold damping with many resonator modes in the quantum mechanical treatment we start with the equations of motion given by
\begin{subequations}
\begin{align}
\label{App.Full1}
 \dot{q}_j &= \omega_j   p_j,\\
 \dot{p}_j &= -\omega_j   q_j - \gamma_j   p_j + G_j   x + \xi_j - \int_{-\infty}^{\infty}ds g_j(t-s)  y^{\text{est}}(s),\\
 \dot{x} &= -\kappa   x + \sqrt{2\kappa}x^{\text{in}}, \\
 \dot{y} &= -\kappa   y + \textstyle \sum_{j=1}^{N} G_j  q_j + \sqrt{2\kappa}y^{\text{in}},
\end{align}
\end{subequations}
where the effective cavity detuning is kept at zero $\Delta = 0$.

\subsection*{Feedback details}
The quadrature component that is injected into the feedback mechanism $  y^{\text{est}}$ is the estimated intracavity phase quadrature. This results from a measurement of the output quadrature $y^{\text{out}} = \sqrt{2\kappa} y(t) - y^{\text{in}}(t)$ additionally considering a detector with quantum efficiency $\eta$ which is modeled by an ideal detector preceded by a beam splitter with transmissivity $\sqrt{\eta}$, which mixes the input field with an uncorrelated vacuum field $y^v(t)$. The estimated phase quadrature is decribed by
\begin{subequations}
\begin{align}
\label{App.Full2}
y^{\text{est}}(t) &= y(t) - \frac{y^{\text{in}}(t) + \sqrt{\eta^{-1} -1}y^v(t)}{\sqrt{2\kappa}}.
\end{align}
\end{subequations}

\subsection*{Eliminating the cavity quadratures}

We can eliminate the cavity field quadratures by formally integrating their equations of motion to obtain:
\begin{subequations}
\begin{align}
x(t) &= G_j \int^{t}_{-\infty} ds e^{-\kappa(t-s)}x^{\text{in}}(s),\\
y(t) &= \int^{t}_{-\infty}ds e^{-\kappa (t-s)}\sum_{j=1}^{N}G_j q_{j}(s)+\sqrt{2\kappa}\int^{t}_{-\infty}ds e^{-\kappa (t-s)} y^{\text{in}}(s).
\end{align}
\end{subequations}
The $y^{\text{est}}(s)$ term introduces both terms proportional to the $q_j$ as well as noise terms stemming from the cavity input noise $y^{\text{in}}(s)$ as well as from the vacuum filled port noise $y^v$. We can first work out the terms coming from $y$ as:
\begin{subequations}
\begin{align}
\label{App.Eq1}
\nonumber
(g_j \ast y) &= g^{(j)}_{\text{cd}} \omfb  \int_{-\infty}^{\infty}ds e^{-\omfb (t-s)}\delta(t-s) y(s) - g^{(j)}_{\text{cd}}\omfb ^{2}\int_{-\infty}^{\infty}ds \theta(t-s)e^{-\omfb (t-s)} y(s) \\
& = \int_{-\infty}^{t} ds \frac{\kappa e^{-\kappa(t-s)} - \omfb e^{-\omfb (t-s)}}{(\kappa - \omfb )} \left[\textstyle \sum_{k=1}^{N} g^{(j)}_{\text{cd}}\omfb G_k q_{k}(s) + \sqrt{2\kappa}g^{(j)}_{\text{cd}}\omfb y^{\text{in}}(s)\right].
\end{align}
\end{subequations}
To obtain a dependence with respect to $p_j$ we apply integration by parts noticing that the convolution above contains a derivative of the following function:
\begin{equation}
h(t-s)=\frac{e^{-\kappa(t-s)}-e^{-\omfb (t-s)}}{\omfb -\kappa}
\end{equation}
and the relation $\dot{q}_j = \omega_j p_j$ and we obtain
\begin{align}
\label{App.Eq2}
(g_j \ast y) = \sum_{k=1}^{N} g^{(j)}_{\text{cd}}\omfb G_k \omega_k \int_{-\infty}^{t}ds h(t-s) p_k(s)+ \sqrt{2\kappa}g^{(j)}_{\text{cd}}\omfb  \int_{-\infty}^{t} ds \partial_sh(t-s) y^{\text{in}}(s)
\end{align}
We can now write in simplified notation the reduced set of equations of motion for the $2N$ resonator modes quadratures
\begin{subequations}
\begin{align}
\label{App.Full4}
\dot{q}_j &= \omega_j p_j, \\\nonumber
\dot{p}_j &= -\omega_j q_j - \int^{\infty}_{-\infty}ds\left(\gamma_j\delta(t-s) + g^{(j)}_{\text{cd}}\omfb G_j\omega_j \theta(t-s)h(t-s)\right) p_j(s) - \sum_{k \neq j} g^{(j)}_{\text{cd}}\omfb G_k\omega_k\int^{t}_{-\infty} ds  h(t-s) p_k(s)\\
& + \xi_j+\xi_\text{fb}+ \xi_\text{vac}+ \xi_\text{rp}.
\end{align}
\end{subequations}
The three sources of noise are owed to the direct feedback action, to the feedback filtered vacuum action in the loss port and to the intra-cavity radiation pressure effect:
\begin{subequations}
\begin{align}
\label{App.Full5}
\xi_\text{fb} &= - \frac{g^{(j)}_{\text{cd}}\omfb }{\sqrt{2\kappa}}\int^{\infty}_{-\infty}ds \phi_1(t-s)y^{\text{in}}(s), \\
\xi_\text{vac} &= - \frac{g^{(j)}_{\text{cd}}\omfb }{\sqrt{2\kappa}}\sqrt{\eta^{-1}-1}\int^{\infty}_{-\infty} ds \phi_2(t-s)y^v(s),\\
\xi_\text{rp} &= \sqrt{2\kappa}G_j\int^{\infty}_{-\infty}ds \phi_3(t-s)x^{\text{in}}(s),
\end{align}
\end{subequations}
with the following definitions
\begin{subequations}
\begin{align}
\label{App.Full5}
\phi_{1}(t) &= \theta(t)(\omfb (\omfb +\kappa)e^{-\omfb t}-2\kappa^2 e^{-\kappa t})/(\omfb -\kappa) - \delta(t), \\
\phi_{2}(t) &= \theta(t)\omfb e^{-\omfb t} -\delta(t),\\
\phi_{3}(t) &= \theta(t)e^{-\kappa t},
\end{align}
\end{subequations}
for the convolution kernels.

\subsection*{Performing the fast-feedback fast-cavity approximation}

In the limit of lossy cavity and fast feedback where $\kappa, \omfb  \gg \omega_j$ we can estimate
\begin{equation}
\int^{t}_{-\infty}ds h(t-s) p_j(s) \approx  p_j(t)\int^{t}_{-\infty}ds h(t-s) =\frac{p_j(t)}{\omfb \kappa}
\end{equation}
and we end up with a set of coupled linear differential equations for the mechanical mode quadratures:
\begin{subequations}
\begin{align}
\label{App.Full5}
\dot{q}_j &= \omega_j p_j, \\
\dot{p}_j &= -\omega_j q_j - \Gamma_{jj} p_j - \sum_{k \neq j}\Gamma_{jk} p_k + \xi_j + \xi_\text{fb}+ \xi_\text{vac}+ \xi_\text{rp}.
\end{align}
\end{subequations}
The off-diagonal rates are defined as
\begin{equation}
\Gamma_{jk} = (g^{(j)}_{\text{cd}} G_k\omega_k)/\kappa
\end{equation}
while the diagonal damping rates are the expected independent cooling rates
\begin{equation}
\Gamma_{jj} = \gamma_j+(g^{(j)}_{\text{cd}} G_j\omega_j)/\kappa
\end{equation}

\subsection*{Solving the Lyapunov equation}
The set of differential equations presented in Eq.~\ref{App.Full5} can be cast into the form
\begin{subequations}
\begin{align}
\label{App.Lya1}
\dot{\bold{v}} &= M \bold{v} + \bold{n}_{\text{in}}
\end{align}
\end{subequations}
with $\bold{v} = (\delta q_1, \delta p_1, \dots \delta q_N, \delta p_N)^{\top}$ and $ \bold{n}_{\text{in}} = (0, \tilde{\eta}_1, \dots , 0, \tilde{\eta}_N)$ where all noise terms have been gathered into a single term $\tilde{\eta}_j = \xi_j+\xi_\text{fb}+ \xi_\text{vac}+ \xi_\text{rp}$. From the general solution \begin{equation}
\bold{v}(t) = e^{M(t-t_0)}\bold{v}(t_0) + \int^{t}_{t_0} ds e^{M(t-s)}\bold{n}_{\text{in}}(s)
\end{equation}
we obtain the correlation matrix
\begin{subequations}
\begin{align}
\label{App.Lya2}
V = \langle \bold{v}(t) \bold{v}^{\top}(t) \rangle = \int^{t}_{t_0}ds \int^{t}_{t_0}ds' e^{M(t-s)}\langle \bold{n}_{\text{in}}(s) \bold{n}_{\text{in}}^{\top}(s')\rangle e^{M^{\top}(t-s')},
\end{align}
\end{subequations}
where we have ignored the transient solution which will decay strongly for large times $t$. Regarding the noise correlation term $\langle \bold{n}_{\text{in}}(s)\bold{n}_{\text{in}}^{\top}(s')\rangle$ component wise we obtain $\langle n_{\text{in},i}(s)n_{\text{in},j}(s') \rangle \neq 0$ if $i$ and $j$ are both even numbers, which is resulting in
\begin{subequations}
\begin{align}
\label{App.Lya3}
\nonumber
\langle n_{\text{in},2i}(s)n_{\text{in},2j}(s') \rangle &= \langle \tilde{\eta}_{i}(s)\tilde{\eta}_{j}(s')\rangle \\\nonumber
&= \langle \xi_i(s)\xi_j(s')\rangle + \frac{g^{(i)}_{\text{cd}}g^{(j)}_{\text{cd}}\omfb ^{2}}{2\kappa}\left[ \langle (\phi_1 \ast y^{\text{in}})(s)(\phi_1 \ast y^{\text{in}})(s') \rangle + (\eta^{-1} -1) \langle (\phi_2 \ast y^{v})(s)(\phi_2 \ast y^{v})(s') \rangle\right]\\\nonumber
& + 2\kappa G_i G_j \langle (\phi_3 \ast x^{\text{in}})(s)(\phi_3 \ast x^{\text{in}})(s')\rangle - g^{(i)}_{\text{cd}}\omega_{\text{fb}}G_j \langle (\phi_1 \ast y^{\text{in}})(s)(\phi_3 \ast x^{\text{in}})(s') \rangle  \\\nonumber
& - g^{(j)}_{\text{cd}}\omega_{\text{fb}}G_i \langle (\phi_3 \ast x^{\text{in}})(s)(\phi_1 \ast y^{\text{in}})(s') \rangle \\\nonumber
&= (2\bar{n}_{i}+1)\gamma_{i}\delta_{ij}\delta(s-s') + \frac{g^{(i)}_{\text{cd}}g^{(j)}_{\text{cd}}\omfb ^{2}}{4\kappa\eta}\left(\delta(s-s') - \frac{\omfb }{2}e^{-\omfb |s-s'|} \right) + \frac{G_i G_j}{\kappa}\frac{\kappa}{2}e^{-\kappa|s-s'|}\\
& + i\left(\frac{\omega_{\text{fb}}e^{-\omega_{\text{fb}}|s-s'|}-\kappa e^{-\kappa|s-s'|}}{2(\omega_{\text{fb}}-\kappa)} \right)\omega_{\text{fb}}\left(g^{(i)}_{\text{cd}}G_j \theta(s-s') - g^{(j)}_{\text{cd}}G_i \theta(s'-s)\right),
\end{align}
\end{subequations}
for $i,j \in \{1,\dots,N\}$. For $\omfb , \kappa \gg \omega_j, \Gamma_j$ we can approximate $\delta(t) \approx (\omfb /2)e^{-\omfb  |t|}$ as well as $\delta(t) \approx (\kappa/2)e^{-\kappa |t|}$ resulting in
\begin{subequations}
\begin{align}
\label{App.Lya4}
\langle \tilde{\eta}_{i}(s)\tilde{\eta}_{j}(s')\rangle &\approx \left( (2\bar{n}_{i}+1)\gamma_{i}\delta_{ij} + \frac{G_iG_j}{\kappa}\right)\delta(s-s').
\end{align}
\end{subequations}
For $\delta$-correlated noise we can simplify the correlation matrix to
\begin{subequations}
\begin{align}
\label{App.Lya5}
V &= \int^{t}_{t_0} ds e^{M(t-s)}\mathcal{D}_{\text{in}}e^{M^{\top}(t-s)},
\end{align}
\end{subequations}
where $\mathcal{D}_{\text{in},2i,2j} =  (2\bar{n}_{i}+1)\gamma_{i}\delta_{ij} + G_i G_j/\kappa$ for even index numbers and is zero otherwise.
The Lyapunov equation for the $N$-oscillator system which determines the steady solution of the correlation matrix is given by
\begin{eqnarray}
\label{App.Eq3}
MV + VM^{\top} = -\mathcal{D}_{\text{in}},
\end{eqnarray}
and can be solved exactly. Evaluating the individual components we obtain the set of equations
\begin{subequations}
\begin{align}
\label{App.Eq4}
Y_{ii} &= 0, \\
\omega_j Y_{ij} +\omega_i Y_{ji} &= 0,\\
\Gamma_{ii} X_{ii} + \sum_{j\neq i}\Gamma_{ij} X_{ij} - (2\bar{n}_i + 1)\gamma_i - \frac{G_i^2}{\kappa} &= 0, \\
\omega_i\left(X_{ii} - Z_{ii} \right) - \sum_{j \neq i} \Gamma_{ij} Y_{ij} &= 0, \\
(\omega_i^2 - \omega_j^2)X_{ij} - \frac{\left(\omega_i^2\Gamma_{jj} + \omega_j^2\Gamma_{ii} \right)}{\omega_i}Y_{ij} - \sum_{k \neq i,j} \left(\omega_i \Gamma_{jk}Y_{ik} - \omega_j \Gamma_{ik} Y_{jk} \right) &= 0, \\
\label{App.Eq4a1}
-\frac{\left(\omega_i^2 - \omega_j^2 \right)}{\omega_i} Y_{ij} -\left( \Gamma_{ii} + \Gamma_{jj}\right)X_{ij} -\Gamma_{ji}X_{ii}-\Gamma_{ij}X_{jj} - \sum_{k \neq i,j} \left(\Gamma_{ik}X_{jk} + \Gamma_{jk}X_{ik} \right) + \frac{2G_{i}G_{j}}{\kappa}  &= 0,
\end{align}
\end{subequations}
with $X_{ij} = \langle  p_i p_j + p_j p_i \rangle$, $Y_{ij} = \langle q_i p_j + p_j q_i  \rangle$ and $Z_{ij} = \langle q_i q_j + q_j q_i \rangle$. In the case that $\gamma_j \ll (g^{(j)}_{\text{cd}}G_{j}\omega_{j})/\kappa$ where $\Gamma_{jj} \approx (g^{(j)}_{\text{cd}}G_{j}\omega_{j})/\kappa$ and $\Gamma_{ij} = (g^{(i)}_{\text{cd}}/g^{(j)}_{\text{cd}})\Gamma_{jj}$, we can simplify the expression in Eq.~\ref{App.Eq4a1} and we obtain
\begin{subequations}
\begin{align}
\label{App.Eq4aa}
-\frac{\left(\omega_i^2 - \omega_j^2 \right)}{\omega_i} Y_{ij} - \left(\frac{g^{(j)}_{\text{cd}}}{g^{(i)}_{\text{cd}}}\right)(2\bar{n}_i + 1)\gamma_i - \left(\frac{g^{(i)}_{\text{cd}}}{g^{(j)}_{\text{cd}}}\right)(2\bar{n}_j + 1)\gamma_j - \frac{\left(g^{(j)}_{\text{cd}}G_i - g^{(i)}_{\text{cd}}G_j \right)^2}{\kappa g^{(i)}_{\text{cd}}g_{cd,j}} &= 0.
\end{align}
\end{subequations}

Here, we define
\begin{subequations}
\begin{align}
\label{App.Eq4a}
\Lambda_{ij} &:= \left(\left(\frac{g^{(j)}_{\text{cd}}}{g^{(i)}_{\text{cd}}}\right)(2\bar{n}_i + 1)\gamma_i + \left(\frac{g^{(i)}_{\text{cd}}}{g^{(j)}_{\text{cd}}}\right)(2\bar{n}_j + 1)\gamma_j + \frac{(g^{(j)}_{\text{cd}}G_i - g^{(i)}_{\text{cd}}G_j)^2}{\kappa g^{(i)}_{\text{cd}}g^{(j)}_{\text{cd}}}\right),
\end{align}
\end{subequations}
and we obtain
\begin{subequations}
\begin{align}
\label{App.Eq5}
\langle p_i^2\rangle &= \left(\bar{n}_i + \frac{1}{2} \right)\frac{\gamma_i}{\Gamma_{ii}} + \frac{G^2_i}{2\Gamma_{ii}\kappa} + \sum_{j \neq i}\frac{\Gamma_{ij}}{2\Gamma_{ii}}\left\{ \frac{\left(\omega_i^2\Gamma_{jj} + \omega_j^2\Gamma_{ii} \right)\Lambda_{ij}}{\left(\omega_i^2 -\omega_j^2 \right)^2} + \sum_{k \neq i,j} \frac{1}{\left( \omega_i^2 -\omega_j^2\right)}\left(\frac{\omega_i^2 \Gamma_{jk} \Lambda_{ik}}{\left(\omega_i^2 -\omega_k^2 \right)} - \frac{\omega_j^2 \Gamma_{ik} \Lambda_{jk}}{\left(\omega_j^2 - \omega_k^2 \right)} \right) \right\}, \\
\langle q_i^2\rangle &= \langle p_i^2\rangle + \sum_{j \neq i} \frac{\Gamma_{ij}\Lambda_{ij}}{2\left(\omega_i^2 - \omega_j^2 \right)}.
\end{align}
\end{subequations}
Therefore, the energy of the $j$'s mode is given by
\begin{subequations}
\begin{align}
\label{App.Eq6}
\nonumber
\frac{1}{2}\left(\langle p_i^2\rangle + \langle q_i^2\rangle\right) &= \left(\bar{n}_i + \frac{1}{2} \right)\frac{\gamma_i}{\Gamma_{ii}} + \frac{G_i^2}{2\Gamma_{ii}\kappa} \\
&+ \sum_{j \neq i}\left[ \frac{\Gamma_{ij}}{2\Gamma_{ii}}\left\{\frac{\left(\omega_i^2\Gamma_{jj} + \omega_j^2\Gamma_{ii} \right)\Lambda_{ij}}{\left(\omega_i^2 -\omega_j^2 \right)^2} + \sum_{k \neq i,j} \frac{1}{\left( \omega_i^2 -\omega_j^2\right)}\left(\frac{\omega_i^2 \Gamma_{jk} \Lambda_{ik}}{\left(\omega_i^2 -\omega_k^2 \right)} - \frac{\omega_j^2 \Gamma_{ik}\Lambda_{jk}}{\left(\omega_j^2 - \omega_k^2 \right)} \right) \right\} + \frac{\Gamma_{ij} \Lambda_{ij}}{4\left(\omega_i^2 - \omega_j^2 \right)} \right] \\\nonumber
&\approx \bar{n}_i\frac{\gamma_i}{\Gamma_{ii}} + \frac{G_i^2}{2\Gamma_{ii}\kappa} + \sum_{j \neq i} \left[ \frac{\Gamma_{ij}}{\Gamma_{ii}}\left\{ \frac{\left(\omega_i^2\Gamma_{jj} + \omega_j^2\Gamma_{ii} \right)\left(\left(g^{(j)}_{\text{cd}}\right)^{2}\bar{n}_i\gamma_i + \left(g^{(i)}_{\text{cd}}\right)^{2}\bar{n}_j \gamma_j\right)}{(g^{(i)}_{\text{cd}}g^{(j)}_{\text{cd}})\left(\omega_i^2 -\omega_j^2 \right)^2} \right. \right. \\\nonumber
&+ \left. \sum_{k \neq i,j} \frac{1}{\left( \omega_i^2 -\omega_j^2\right)}\left(\frac{\omega_i^2\Gamma_{jk}\left(\left( g^{(k)}_{\text{cd}}\right)^{2}\bar{n}_i\gamma_i + \left(g^{(i)}_{\text{cd}}\right)^{2}\bar{n}_k\gamma_k\right)}{(g^{(i)}_{\text{cd}}g^{(k)}_{\text{cd}})\left(\omega_i^2 -\omega_k^2 \right)} - \frac{\omega_j^2 \Gamma_{ik} \left( \left(g^{(k)}_{\text{cd}}\right)^{2}\bar{n}_j\gamma_j + \left(g^{(j)}_{\text{cd}}\right)^{2}\bar{n}_k\gamma_k \right)}{(g^{(j)}_{\text{cd}}g^{(k)}_{\text{cd}})\left(\omega_j^2 - \omega_k^2 \right)} \right) \right\} \\
& + \left. \frac{\Gamma_{ij}\left(\left(g^{(j)}_{\text{cd}}\right)^{2}\bar{n}_i \gamma_i + \left(g^{(i)}_{\text{cd}}\right)^{2}\bar{n}_j \gamma_j\right)}{2(g^{(i)}_{\text{cd}}g^{(j)}_{\text{cd}})\left(\omega_i^2 - \omega_j^2 \right)} \right] .
\end{align}
\end{subequations}
The term $G_i^2/(2\Gamma_{ii}\kappa) = G_i/\left( 2g^{(i)}_{\text{cd}}\omfb \omega_{i}\right)$ is in general smaller than one and can be mostly ignored if $\bar{n}_i\gamma_i/\Gamma_{ii} > 1$.
For a sequence of frequencies with $\omega_j \approx \omega + j\Delta \omega$ and $\omega \gg \Delta \omega$ we obtain for $\omega_i^2 - \omega_j^2 \approx 2\omega(i-j)\Delta \omega $. By considering only nearest neighbors in frequencies since the terms decay quadratically with distance we obtain
\begin{subequations}
\begin{align}
\label{App.Eq7}
\frac{1}{2}\left(\langle p_i^2\rangle + \langle q_i^2\rangle\right) &\approx \bar{n}_i\frac{\gamma_i}{\Gamma_{ii}} + \frac{1}{4\Delta \omega^2}\sum_{\begin{smallmatrix} \langle i,j \rangle \\ j \neq i \end{smallmatrix}}\Gamma_{ij}\left\{ \left(1 + \frac{\Gamma_{jj}}{\Gamma_{ii}}\right)(\bar{n}_i\gamma_i + \bar{n}_j \gamma_j) + \sum_{\begin{smallmatrix} \langle i,j,k \rangle \\ k \neq i,j \end{smallmatrix}} \left(\frac{\Gamma_{jk}(\bar{n}_i\gamma_i + \bar{n}_k \gamma_k)}{\Gamma_{ii}(i-j)(i-k)}-\frac{\Gamma_{ik}(\bar{n}_j\gamma_j + \bar{n}_k \gamma_k)}{\Gamma_{ii}(i-j)(j-k)} \right)\right\},
\end{align}
\end{subequations}
showing that the lower bound of the energy $\bar{n}_i \gamma_i/\Gamma_{ii}$ can be reached when $\Delta \omega$ is much larger than $\Gamma_{ij}$. Here, we have considered that $g^{(i)}_{\text{cd}} \approx g_{cd}$ for all $i \in \{1, \dots, N \}$.

\newpage
\section{Cooling of two adjacent modes}
\label{C}

\begin{figure}[t]
\includegraphics[width=1.0\columnwidth]{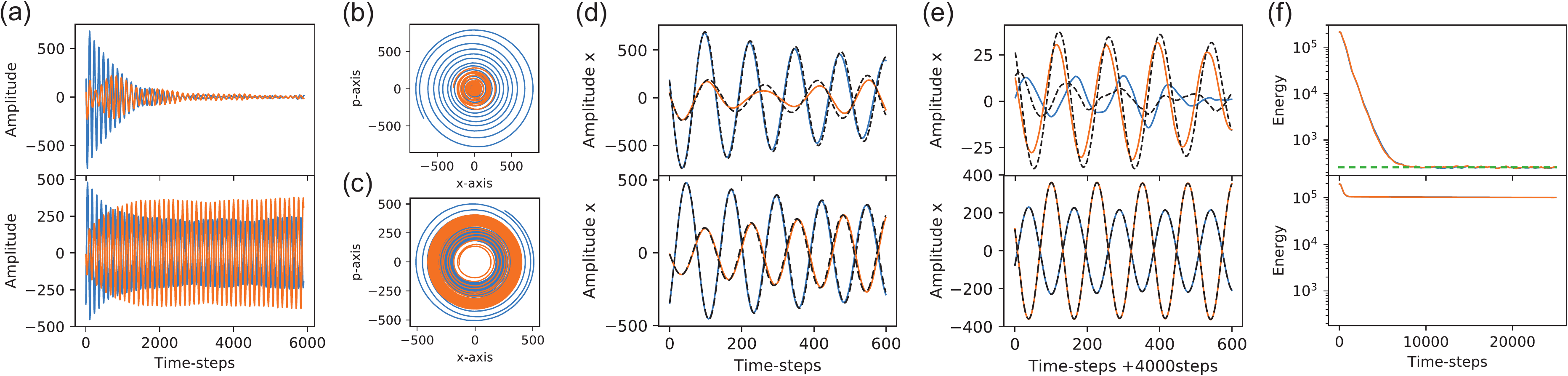}
\caption{\emph{Cold damping for two modes}. (a) Full solutions for cold damping using convolutions for two modes with $\omega_{1,2} = (1, 0.9)$ and in the degenerate case with $\omega_{1,2} = (1,1)$. In (b) and (c) the corresponding phase space trajectories are presented. The magnified signal presented in (d) and (e) show a comparison between the full solutions (solid lines) and the ones with approximated damping rates (dashed lines) at the beginning and a later stage of the evolution, respectively. In the degenerate case the effect of the feedback stops when both modes have a relative phase shift of $\pi$. In (f) the average energy is presented as a function of time. The dashed green line shows the final energy at steady state obtained by the Lyapunov equation for two oscillators. The simulation parameters are given by $\gamma_{1,2} = (4,3)\times 10^{-5}\omega_1$, $G_{1,2} = (0.16, 0.1)\times \omega_1$, $\kappa = 3\omega_1$, $\omfb  = 3.5\omega_1$, $g_{cd,1,2} = (0.8, 0.8)$ and $\tau = 0.05\omega^{-1}$.}
\label{Appfig1}
\end{figure}

To visualize the feedback cooling process we perform classical simulations of the stochastic differential equations.
In Fig.~\ref{Appfig1} we show the results for cooling two modes close to and at frequency degeneracy. The simulation is performed  using the full convolutional description of the feedback process (solid lines) and is compared to the approximated form (dashed lines), which show good agreement. Especially in the degenerate case at a large time duration from initialization as presented in Fig.~\ref{Appfig1}e it is visible that the feedback stops when both modes acquire a phase shift of $\pi$ with respect to each other. Fig.~\ref{Appfig1}f shows the average energy over many trajectories of the two mode system. In the case of frequency degeneracy only up to half of the initial energy of the system is removed since only the bright mode can be accessed by the method.

\section{Numerical integration of Langevin equations}
\label{D}

To test the results derived by solving the Lyapunov equation, we perform numerical Monte-Carlo simulations for the equations of motion. Here, the initial conditions are obtained from a Boltzmann distribution representing the initial thermal state. The numerical integration can be obtained from the differential form of the stochastic differential equations of motion
\begin{subequations}
\begin{align}
\label{App.Eq8}
dq_j &= \omega_j p_j dt, \\
dp_j &= -\omega_j q_j dt - \gamma_j p_j dt - (g_j\ast y) dt + \sqrt{(2\bar{n}_j+1)\gamma_j}dW(t),\\
dy &= -\kappa y dt + \sum_{j=1}^{N} G_j q_j dt,
\end{align}
\end{subequations}
where $dW(t)$ describes an infinitesimal Wiener increment ($dW^2 = dt$) that guarantees that the fluctuation dissipation theorem is fulfilled \cite{Jacobs2010Stochastic}. In our case we use the Runge-Kutta fourth-order method (RK4) that guarantees numerical stability for the integration.

\section{Numerical simulations for many modes}
\label{E}
\begin{figure}[t]
\includegraphics[width=1.0\columnwidth]{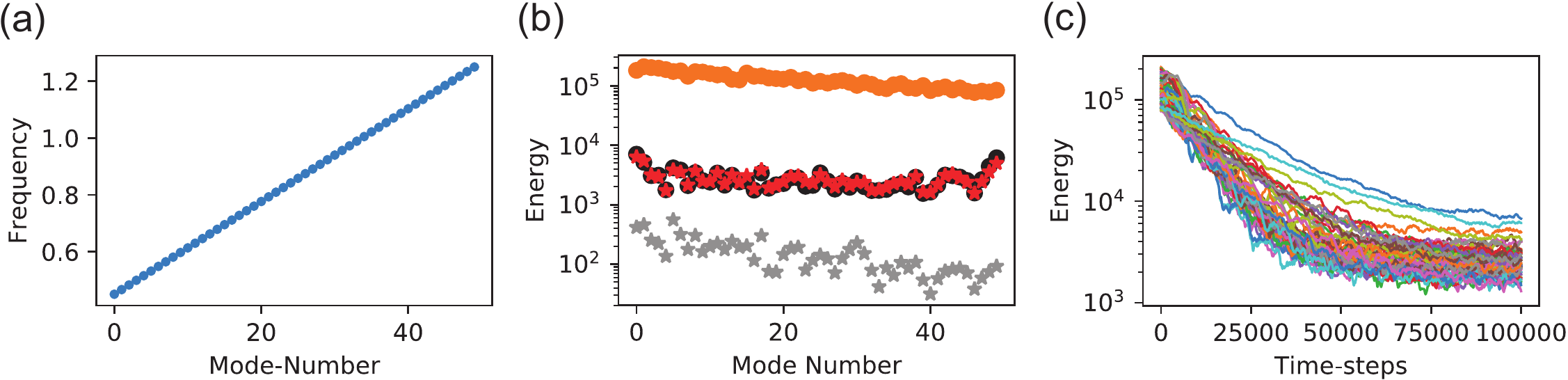}
\caption{\emph{Cold damping of many modes }. (a) Linear dispersion for $N=50$ oscillator modes. (b) Initial (orange disks) and steady state (black disks) occupations of the $50$-modes obtained from the dynamical simulations are presented and compared to the results from the Lyapunov equation (red stars) and in the case of independent damping (gray stars). In (c) the dynamical evolution of the average energies reaching convergence are shown for all modes.}
\label{Appfig2}
\end{figure}

The analytical versus numerical comparison in the main text is based on a small number of resonances such that the results are easily comprehensible and easy to visualize graphically. However, cooling of many more modes is possible using the same technique as we numerically prove in Fig.~\ref{Appfig2}. A limitation for extending numerics to even more modes comes from the limited available computational resources.

\section{Bright and dark mode cooling dynamics}
\label{F}
The cavity output fed into the cold-damping loop contains information solely on a linear combination of individual modes momentum quadratures. Moving to generalized quadratures can provide insight into the efficiency of cooling versus frequency degeneracy (or level of disorder) in the system. Analytical considerations on the simplified system of equations Eqs.~\ref{App.Full5} can already shed light onto the \textit{bright} versus \textit{dark modes} relevance in the cooling process. Let us assume $\Gamma_{jk}=\Gamma$ for set of indexes (we have checked that the predictions of the full equations of motion and of the simplified version of Eqs.~\ref{App.Full5} indeed agree and show the relevant dynamics around degeneracy points). An analysis of the simplified equations:
\begin{subequations}
\begin{align}
\label{CDEq.5}
  \dot{q}_j &= \omega_j   p_j,\\
  \dot{p}_j &= -\omega_j   q_j - (\Gamma+\gamma)  p_j -\Gamma \textstyle \sum_{k\neq j}p_k+ \xi_j+\xi_{\text{opt},j},
\end{align}
\end{subequations}
shows that a generalized momentum quadrature can be defined as $\mathcal{P}_1=1/\sqrt{N}\sum_j p_j$ (with the corresponding position quadrature $\mathcal{Q}_1=1/\sqrt{N}\sum_j q_j$) corresponding to the bright mode. Notice that the commutation relation for these quadratures are fulfilled. Dark modes (not directly addressed by the feedback loop) can be defined via a Gram-Schmidt procedure as $\mathcal{P}_k=\sum_j \alpha_{kj}p_j$ and $\mathcal{Q}_k=\sum_j \alpha_{kj}q_j$ such that commutations are insured. Notice also that orthogonality to each other and to the bright mode is fulfilled meaning that $\sum_j \alpha_{jk}=0$ and $\sum_j \alpha^*_{jk} \alpha_{jk'}=\delta_{kk'}$. Conservation of energy requires that $\sum_j \omega_j (p^2_j+q^2_j)=\sum_k \Omega_k (\mathcal{P}^2_k+\mathcal{Q}^2_k)$ where the generalized quadratures eigen-frequencies depend on the coefficients $\alpha_{jk}$. With notations $\bold{v}=(q_1,p_1,...q_N,p_N)$ and $\bold{u}=(\mathcal{Q}_1,\mathcal{P}_1,...\mathcal{Q}_N,\mathcal{P}_N)$ we can write a general transformation $\bold{u}=T \bold{v}$ and an equation of motion for the vector $\bold{u}$ of generalized quadratures
\begin{align}
\dot{\bold{u}}=T \left( M_{\omega,\gamma}- M_\Gamma \right )T^{\top}+\bold{u}_{\text{in}}.
\end{align}
with
\begin{subequations}
\begin{align}
M_{\omega,\gamma} &= \left( \begin{smallmatrix} 0 & \omega_1 & 0 & 0 & \cdots & 0 & 0 \\ -\omega_1 & -\gamma & 0 & 0 & \cdots & 0 & 0 \\
0 & 0 & 0 & \omega_2 & \cdots & 0 & 0 \\ 0 & 0 & -\omega_2 & -\gamma & \cdots & 0 & 0 \\ \vdots & \vdots & \vdots & \vdots
& \ddots & \vdots & \vdots \\ 0 & 0 & 0 & 0 & \cdots & 0 & \omega_{N} \\ 0 & 0 & 0 & 0 & \cdots & -\omega_{N} & -\gamma \end{smallmatrix} \right) \qquad \text{and} \qquad M_\Gamma = \left( \begin{smallmatrix} 0 & 0 & 0 & 0 & \cdots & 0 & 0 \\ 0 & -\Gamma & 0 & -\Gamma & \cdots & 0 & -\Gamma \\
0 & 0 & 0 & 0 & \cdots & 0 & 0 \\ 0 & -\Gamma & 0 & -\Gamma & \cdots & 0 & -\Gamma \\ \vdots & \vdots & \vdots & \vdots
& \ddots & \vdots & \vdots \\ 0 & 0 & 0 & 0 & \cdots & 0 & 0 \\ 0 & -\Gamma & 0 & -\Gamma & \cdots & 0 & -\Gamma \end{smallmatrix} \right),
\end{align}
\end{subequations}
The transformation is defined such that it diagonalizes the damping matrix and can be written as
\begin{subequations}
\begin{align}
T &= \left( \begin{smallmatrix} \alpha_{11} & 0 & \alpha_{12} & 0 & \cdots & \alpha_{1N} & 0 \\ 0 & \alpha_{11} & 0 & \alpha_{12} & \cdots & 0 & \alpha_{1N} \\
\alpha_{21} & 0 & \alpha_{22} & 0 & \cdots & \alpha_{2N} & 0 \\ 0 & \alpha_{21} & 0 & \alpha_{22} & \cdots & 0 & \alpha_{2N} \\ \vdots & \vdots & \vdots & \vdots
& \ddots & \vdots & \vdots \\ \alpha_{N1} & 0 & \alpha_{N2} & 0 & \cdots & \alpha_{NN} & 0 \\ 0 & \alpha_{N1} & 0 & \alpha_{N2} & \cdots & 0 & \alpha_{NN} \end{smallmatrix} \right),
\end{align}
\end{subequations}
where the vector $(\alpha_{11}, \alpha_{12}, \cdots, \alpha_{1N}) = 1/\sqrt{N}(1,1,\cdots,1)$ describes the bright symmetric mode.
The diagonalized matrix has the form
\begin{subequations}
\begin{align}
T M_{\Gamma}T^{\top} &= \left( \begin{matrix} 0 & 0 & 0 & 0 & \cdots & 0 & 0 \\ 0 & -N\Gamma & 0 & 0 & \cdots & 0 & 0 \\
0 & 0 & 0 & 0 & \cdots & 0 & 0 \\ 0 & 0 & 0 & 0 & \cdots & 0 & 0 \\ \vdots & \vdots & \vdots & \vdots
& \ddots & \vdots & \vdots \\ 0 & 0 & 0 & 0 & \cdots & 0 & 0 \\ 0 & 0 & 0 & 0 & \cdots & 0 & 0 \end{matrix} \right).
\end{align}
\end{subequations}
showing that feedback damping only affects the bright mode. However, sympathetic cooling of all other collective modes appears from off-diagonal coupling terms of bright to dark modes in the transformed  matrix
\begin{subequations}
\begin{align}
T M_{\omega,\gamma}T^{\top} &= \left( \begin{smallmatrix} 0 & \sum \alpha_{1j}^2\omega_j & 0 & \sum \alpha_{1j}\alpha_{2j}\omega_j & \cdots & 0 & \sum \alpha_{1j}\alpha_{Nj}\omega_j \\ -\sum \alpha_{1j}^2\omega_j & 0 & -\sum \alpha_{1j}\alpha_{2j}\omega_j & 0 & \cdots & -\sum \alpha_{1j}\alpha_{Nj}\omega_j & 0 \\
0 & \sum \alpha_{2j}\alpha_{1j}\omega_j & 0 & \sum \alpha_{2j}^2\omega_j & \cdots & 0 & \sum \alpha_{2j}\alpha_{Nj}\omega_j \\ -\sum \alpha_{2j}\alpha_{1j}\omega_j & 0 & -\sum \alpha_{2j}^2\omega_j & 0 & \cdots & -\sum \alpha_{2j}\alpha_{Nj}\omega_j & 0 \\ \vdots & \vdots & \vdots & \vdots
& \ddots & \vdots & \vdots \\ 0 & \sum \alpha_{Nj}\alpha_{1j}\omega_j & 0 & \sum \alpha_{Nj}\alpha_{2j}\omega_j & \cdots & 0 & \sum \alpha_{Nj}^2\omega_j \\ -\sum \alpha_{Nj}\alpha_{1j}\omega_j & 0 & -\sum \alpha_{Nj}\alpha_{2j}\omega_j & 0 & \cdots & -\sum \alpha_{Nj}^2\omega_j & 0 \end{smallmatrix} \right)
\end{align}
\end{subequations}
The frequencies of the generalized quadratures can be determined as $\Omega_k=\sum_j \alpha_{kj}^2\omega_j$ and the intermode couplings are defined as $g_{kk'}=\sum_j \alpha_{kj}\alpha_{k'j}\omega_j$. The bright mode always exhibits a frequency as a harmonic average of all individual mode frequencies: $\Omega_1=\sum_j \omega_j/N$. In these simplified notations one can then write the equations of motion:
\begin{subequations}
\begin{align}
	\dot{\mathcal{Q}}_1 &= \Omega_1 \mathcal{P}_1 + \sum_{k'=2}^{N}g_{1k'}\mathcal{P}_k',\\
	\dot{\mathcal{P}}_1 &= -\Omega_1 \mathcal{Q}_1 - (N\Gamma+\gamma)\mathcal{P}_1 -\sum_{k'=2}^{N}g_{1k'}\mathcal{Q}_k'  + \Xi_k,\\
    \dot{\mathcal{Q}}_k &= \Omega_k   \mathcal{P}_k + \sum_{k'=2}^{N}g_{kk'}\mathcal{P}_k' ,\\
	\dot{\mathcal{P}}_k &= -\Omega_k   \mathcal{Q}_k - \gamma \mathcal{P}_k - \sum_{k'=2}^{N}g_{kk'}\mathcal{P}_k' + \Xi_k.
\end{align}
\end{subequations}
Note that only the bright mode is directly driven by the feedback and exhibits an enhanced damping rate $N\Gamma$. In the next step, the coupling of the dark modes to the damped bright mode stemming from off-diagonal terms proportional to frequency mismatches $g_{kk'}$ can then lead to sympathetic cooling of all degrees of freedom.

\subsection{The fully degenerate case}
In the case of full degeneracy where $\omega_i = \omega$ for all $i\in \{1, \dots, N\}$ the intermode couplings vanish altogether and the equations of motion for collective modes show full separation:
\begin{subequations}
\begin{align}
	\dot{\mathcal{Q}}_1 &= \omega \mathcal{P}_1,\\
	\dot{\mathcal{P}}_1 &= -\omega \mathcal{Q}_1 - (N\Gamma+\gamma)\mathcal{P}_1 + \Xi_j,\\
    \dot{\mathcal{Q}}_j &= \omega   \mathcal{P}_j,\\
	\dot{\mathcal{P}}_j &= -\omega   \mathcal{Q}_j - \gamma \mathcal{P}_j + \Xi_j.
\end{align}
\end{subequations}
The bright mode the is cooled at $N\Gamma$ while all the dark modes are left completely unaffected (as there is no damping term except for the $\gamma$ associated with the thermalization to the environmental temperature). The efficiency of the cooling process is then minimal as only $1/N$ of the total energy is removed.

\subsection{Illustration: two and three modes}
Let us illustrate the role of frequency disorder in coupling bright to dark modes in the case on only two or three distinct modes present. For two modes the dark combination is defined by the coefficients $(1/\sqrt{2},-1/\sqrt{2})$. The equations of motion become
\begin{subequations}
\begin{align}
	\dot{\mathcal{Q}}_1 &= \frac{\omega_1+\omega_2}{2} \mathcal{P}_1 + \frac{\omega_1-\omega_2}{2} \mathcal{P}_2,\\
	\dot{\mathcal{P}}_1 &= -\frac{\omega_1+\omega_2}{2} \mathcal{Q}_1 - (2\Gamma+\gamma)\mathcal{P}_1 -\frac{\omega_1-\omega_2}{2} \mathcal{Q}_2 + \Xi_1,\\
    \dot{\mathcal{Q}}_2 &= \frac{\omega_1+\omega_2}{2}   \mathcal{P}_2 + \frac{\omega_1-\omega_2}{2} \mathcal{P}_1,\\
	\dot{\mathcal{P}}_2 &= -\frac{\omega_1+\omega_2}{2}   \mathcal{Q}_2 - \gamma \mathcal{P}_2  -\frac{\omega_1-\omega_2}{2} \mathcal{Q}_1 + \Xi_2.
\end{align}
\end{subequations}
Sympathetic cooling of the dark mode is driven by the indirect coupling to the feedback loop via the direct coupling to the position quadrature of the bright mode proportional to the frequency difference.\\
For three modes we obtain a more complicated coupling map where as expected the bright mode is damped quickly at $3\Gamma$
\begin{subequations}
\begin{align}
	\dot{\mathcal{Q}}_1 &= \frac{\omega_1+\omega_2+\omega_3}{3} \mathcal{P}_1 + \frac{\omega_1-2\omega_2+\omega_3}{3\sqrt{2}} \mathcal{P}_2 + \frac{\omega_1-\omega_3}{\sqrt{6}} \mathcal{P}_3,\\
	\dot{\mathcal{P}}_1 &= -\frac{\omega_1+\omega_2+\omega_3}{3} \mathcal{Q}_1 - (3\Gamma+\gamma)\mathcal{P}_1 -\frac{\omega_1-2\omega_2+\omega_3}{3\sqrt{2}} \mathcal{Q}_2 - + \frac{\omega_1-\omega_3}{\sqrt{6}} \mathcal{Q}_3 + \Xi_1,\\
    \dot{\mathcal{Q}}_2 &= \frac{\omega_1+4\omega_2+\omega_3}{6} \mathcal{P}_2 + \frac{\omega_1-2\omega_2+\omega_3}{3\sqrt{2}} \mathcal{P}_1 + \frac{\omega_1-\omega_3}{2\sqrt{3}} \mathcal{P}_3  ,\\
	\dot{\mathcal{P}}_2 &= -\frac{\omega_1+4\omega_2+\omega_3}{6} \mathcal{Q}_2 - \gamma \mathcal{P}_2 -\frac{\omega_1-2\omega_2+\omega_3}{3\sqrt{2}} \mathcal{Q}_1 -\frac{\omega_1-\omega_3}{2\sqrt{3}} \mathcal{Q}_3 + \Xi_2,\\
\dot{\mathcal{Q}}_3 &= \frac{\omega_1+\omega_3}{2}   \mathcal{P}_3 +\frac{\omega_1-\omega_3}{\sqrt{6}} \mathcal{P}_1 + \frac{\omega_1-\omega_3}{2\sqrt{3}}\mathcal{P}_2,\\
	\dot{\mathcal{P}}_3 &= -\frac{\omega_1+\omega_3}{2}  \mathcal{Q}_3 - \gamma \mathcal{P}_3  -\frac{\omega_1-\omega_3}{\sqrt{6}} \mathcal{Q}_1 -\frac{\omega_1-\omega_3}{2\sqrt{3}}\mathcal{Q}_2  + \Xi_2.
\end{align}
\end{subequations}
The bright mode then in turn couples to both dark modes which are also coupled among each other.

\begin{figure}[b]
\includegraphics[width=0.8\columnwidth]{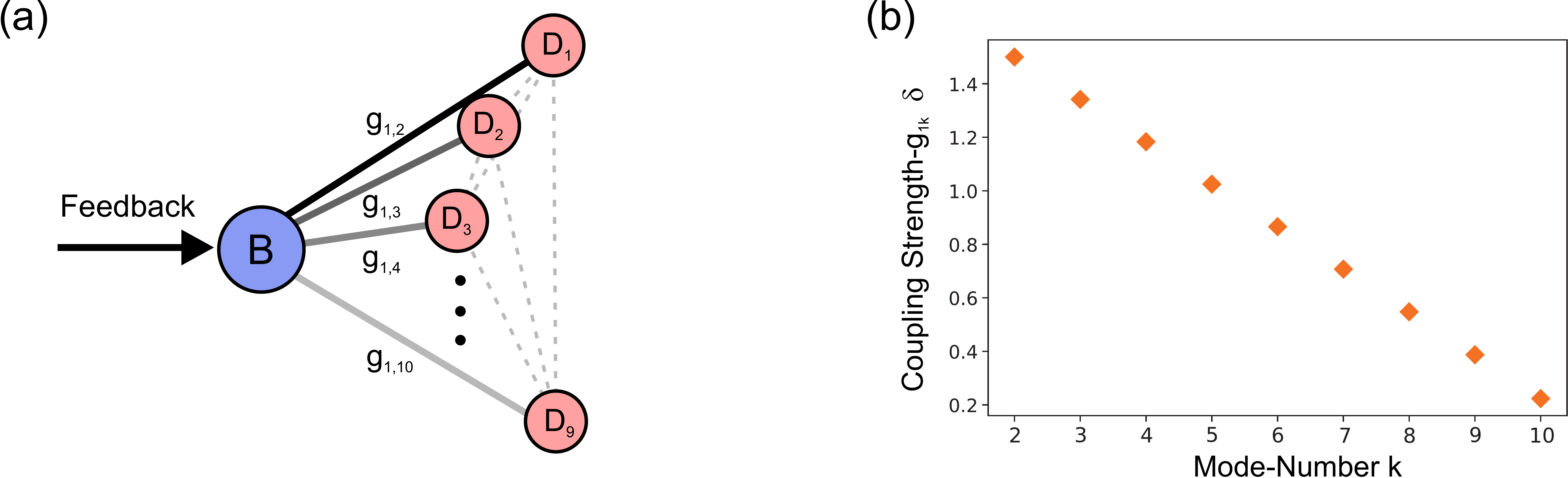}
\caption{\emph{Mode coupling }. (a) The bright mode $B$ experiences directly the action of the feedback loop and sympathetically cools the rest of the modes $D_j$ (depicted is the case of 10 independent modes). (b) Plot of the coupling strength between the bright mode and dark modes (in units of the frequency difference $\delta$) in the linear dispersion case showing a linear decrease.}
\label{Appfig3}
\end{figure}

\subsection{Coupling rates for the linear dispersion case}
For a sequence of frequencies $\omega_{j} = \omega + (j-1)\delta$ we obtain for the matrix elements
\begin{subequations}
\begin{align}
(T M_{\omega,\gamma} T^{\top})_{(2i-1)(2k)} &= \sum_{j=1}^{N}\alpha_{ij}\alpha_{kj}(\omega+(j-1)\delta) = \omega \delta_{ik} + \delta\sum_{j=1}^{N}(j-1)\alpha_{ij}\alpha_{kj} \\
& = (\omega - \delta)\delta_{ik} + \delta\sum_{j=1}^{N}j\alpha_{ij}\alpha_{kj},
\end{align}
\end{subequations}
where the other non zero entries are following $(TMT^{\top})_{(2i)(2k-1)} = -(TMT^{\top})_{(2i-1)(2k)}$.
This results in the following coupling matrix
\begin{subequations}
\begin{align}
T M_{\omega,\gamma}T^{\top} &= \left( \begin{smallmatrix} 0 & \omega +(N-1)\delta/2 & 0 & \delta\sum j\alpha_{1j}\alpha_{2j} & \cdots & 0 & \delta\sum j\alpha_{1j}\alpha_{Nj} \\ -(\omega +(N-1)\delta/2) & 0 & -\delta\sum j\alpha_{1j}\alpha_{2j} & 0 & \cdots & -\delta\sum j\alpha_{1j}\alpha_{Nj} & 0 \\
0 & \delta\sum j\alpha_{2j}\alpha_{1j} & 0 & \omega +(N-1)\delta/2 & \cdots & 0 & \delta\sum j\alpha_{2j}\alpha_{Nj} \\ -\delta\sum j\alpha_{2j}\alpha_{1j} & 0 & -(\omega +(N-1)\delta/2) & 0 & \cdots & -\delta\sum j\alpha_{2j}\alpha_{Nj} & 0 \\ \vdots & \vdots & \vdots & \vdots
& \ddots & \vdots & \vdots \\ 0 & \delta\sum j\alpha_{Nj}\alpha_{1j} & 0 & \delta\sum j\alpha_{Nj}\alpha_{2j} & \cdots & 0 & \omega +(N-1)\delta/2 \\ -\delta\sum j\alpha_{Nj}\alpha_{1j} & 0 & -\delta\sum j\alpha_{Nj}\alpha_{2j} & 0 & \cdots & -(\omega +(N-1)\delta/2) & 0 \end{smallmatrix} \right).
\end{align}
\end{subequations}
The common frequency of all collective modes is $\omega +(N-1)\delta/2$ and the off diagonal elements show the bright-dark as well as the dark-dark couplings. Notice that the bright to dark rates diminish progressively when scanning through al the collective modes. This is illustrated in Fig.~\ref{Appfig3} for a case including 10 independent vibrational resonances.

\end{document}